\documentclass[conference,10pt]{IEEEtran}
\usepackage[utf8]{inputenc}
\usepackage{amsmath,amssymb}
\usepackage{graphicx}
\usepackage{booktabs}
\usepackage{cite}
\usepackage{url}
\usepackage{multirow}
\usepackage{array}
\usepackage[hidelinks]{hyperref}
\usepackage{orcidlink}
\usepackage{tikz}
\usetikzlibrary{shapes.geometric,arrows.meta,positioning,fit,calc}
\usepackage{balance}

\graphicspath{{figures/}}

\title{On the Optimal Co-location of Data Centers\\with Renewable Energy Sources in SIDS}

\author{\IEEEauthorblockN{Shankar Ramharack\,\orcidlink{0000-0003-3759-7333}}
\IEEEauthorblockA{\textit{Engineering Department} \\
\textit{M\&E Partners (Barbados) Limited} \\
St.~Philip, Barbados, W.I.\\
sramharack@ieee.org}
\and
\IEEEauthorblockN{Patrick Hosein\,\orcidlink{0000-0003-1729-559X}}
\IEEEauthorblockA{\textit{Department of Computer Science} \\
\textit{The University of the West Indies}\\
St.~Augustine, Trinidad and Tobago\\
patrick.hosein@uwi.edu}
\and
\IEEEauthorblockN{Rajiv Sahadeo}
\IEEEauthorblockA{Trinidad and Tobago, W.I.\\
rajivsahadeo@outlook.com}
}

\begin{document}
\maketitle

\begin{abstract}
The rapid expansion of data centers presents both economic opportunities and systemic risks for Caribbean Small Island Developing States~(SIDS). This paper develops a PyPSA-based capacity expansion framework for optimal co-location of data centers with renewable energy infrastructure, incorporating behind-the-meter battery storage, workload flexibility (70\%/30\% inflexible/flexible split), and multi-bus transmission topology. Two case studies are examined: Trinidad~\&~Tobago (single-bus, 92\,MW Brechin Castle solar, 5\,MW data center) and Jamaica (4-bus 138\,kV JPS network, 15\,MW data center at Kingston). Jamaica's optimum deploys 405\,MW solar, 278\,MW wind, 800\,MWh grid-scale BESS, and 84\,MWh behind-the-meter BESS, achieving 42.5\% renewable penetration; Trinidad's gas-anchored optimum builds no new RE, and a sensitivity analysis shows solar entering at gas costs above \$65--70/MWh, a carbon price of \$25--50/tCO$_2$, or solar CAPEX of \$800/kW. A post-hoc distributionally robust stress test evaluates hurricane resilience, and a novel 10-factor siting framework addresses the regulatory gap in SIDS that lack data-center-specific zoning. Results show that energy-optimal locations may score poorly on societal criteria---noise, freshwater stress, zoning readiness. This demonstrates the need for integrated planning tools in small island contexts.
\end{abstract}


\section{Introduction}

Global data center electricity consumption reached approximately 460\,TWh in 2024, with projections exceeding 1{,}000\,TWh by 2030 driven by artificial intelligence workloads~\cite{iea2024}. In the United States, over \$64~billion in data center projects have faced delays or cancellations from community opposition over noise, water consumption, and grid strain~\cite{dcwatch2025}. Caribbean SIDS import-dependent economies with fragile power grids and acute hurricane exposure. Consquently, hosting data centers introduces systemic risks that existing planning frameworks do not address.

Jamaica's power system, operated by Jamaica Public Service Company~(JPS), serves approximately 650\,MW peak demand through 400\,km of 138\,kV and 800\,km of 69\,kV transmission connecting 52~substations~\cite{jps_grid}. A 15\,MW data center at Kingston would represent 2.3\% of peak demand. Trinidad~\&~Tobago's 1{,}800\,MW system which is anchored by natural gas from the National Gas Company~(NGC) and recently augmented by the landmark 92\,MW Brechin Castle solar farm~\cite{brechin2025} operates with substantial generation headroom, offering a contrasting context for co-planning analysis.

This paper contributes: (i)~a PyPSA-based co-planning framework with behind-the-meter~(BTM) storage, workload flexibility, and multi-bus transmission based on real JPS topology; (ii)~post-hoc distributionally robust hurricane resilience assessment; and (iii)~a 10-factor siting suitability framework for Caribbean SIDS. Code, data, and results are openly available at \url{https://github.com/sramharack/caribcon26-ocreds}.

\section{Literature Review}

\subsection{The Data Center Energy Problem}

Data centers combine high power density (5--50\,MW per facility), load factors of 0.85--0.95, Tier~III/IV reliability requirements (99.982\%+ uptime), and temperature-dependent cooling loads, which complicates grid integration. Google's 24/7 carbon-free energy~(CFE) initiative demonstrated that hourly matching of data center load to clean generation requires co-optimized storage and flexible scheduling~\cite{google247}. Microsoft's 2024 sustainability report identified workload flexibility as a key enabler, with 20--35\% of training and batch workloads being temporally shiftable~\cite{microsoft2024}.

The siting challenge has intensified. Power, water, and permitting have been identified as the three new pillars of site selection~\cite{dccom2025}. GE~Vernova's consulting framework evaluates grid congestion, climate resilience, and RE resource alignment~\cite{ge_vernova}. In the US, the ACM~COMPASS~2025 study documented that data center cooling systems generate 80--96\,dBA at source, and that Google consumed over 5~billion gallons of water globally in 2023~\cite{compass2025}. Jurisdictions have responded with increasingly specific regulations: 500-foot setbacks, mandatory acoustic modeling, and in some cases outright moratoria~\cite{stafford2025,albemarle2025}.

\subsection{Caribbean SIDS Energy Context}

Caribbean SIDS face electricity costs of \$0.25--0.45/kWh and heavy fossil dependence~\cite{ccreee2023}. Jamaica's RE penetration was 12\% in 2022~\cite{ccreee_jm}, with a government target of 50\% by 2030 supported by IRP-2~\cite{jm_irp2}. Trinidad~\&~Tobago is advancing its energy transition alongside its established gas sector; the Brechin Castle Solar Farm---a joint venture of bp, Shell, and NGC on 186~hectares near Couva---delivered first electrons in July~2025 and contributes approximately 8\% of total generation capacity when fully commissioned~\cite{brechin2025,brechin_bp}. This project demonstrates a replicable model of private-sector-led renewable deployment supported by government facilitation~\cite{brechin_ecp}. No Caribbean nation has data-center-specific zoning frameworks~\cite{uli2024}.

\subsection{Why PyPSA: Open-Source Capacity Expansion}

Commercial tools (PLEXOS, HOMER Pro, Aurora) cost tens of thousands of dollars per license, a material barrier for SIDS ministries, regulators, and universities. Table~\ref{tab:tools} compares relevant frameworks.

\begin{table}[htbp]
\centering
\caption{Energy System Modeling Tools Comparison}
\label{tab:tools}
\setlength{\tabcolsep}{2.5pt}
\begin{tabular}{@{}lcccc@{}}
\toprule
Feature & PyPSA & GenX & PLEXOS & HOMER \\
\midrule
License & Free & Free & Comm. & Comm. \\
Language & Python & Julia & .NET & GUI \\
Multi-bus network & \checkmark & -- & \checkmark & -- \\
Sector coupling & \checkmark & \checkmark & \checkmark & Partial \\
Storage modeling & \checkmark & \checkmark & \checkmark & \checkmark \\
Open solver (HiGHS) & \checkmark & \checkmark & -- & -- \\
Active community & \checkmark & \checkmark & -- & -- \\
PyPSA-Earth (global) & \checkmark & -- & -- & -- \\
\bottomrule
\end{tabular}
\end{table}

PyPSA~\cite{pypsa} was selected because it is MIT-licensed, models multi-bus networks with transmission constraints natively, runs with the open-source HiGHS solver~\cite{highs}, and is backed by an active community and the global PyPSA-Earth model~\cite{pypsa_earth}. Its Python API scales from single-bus to continental networks, covering both case studies here.

\section{Methodology}

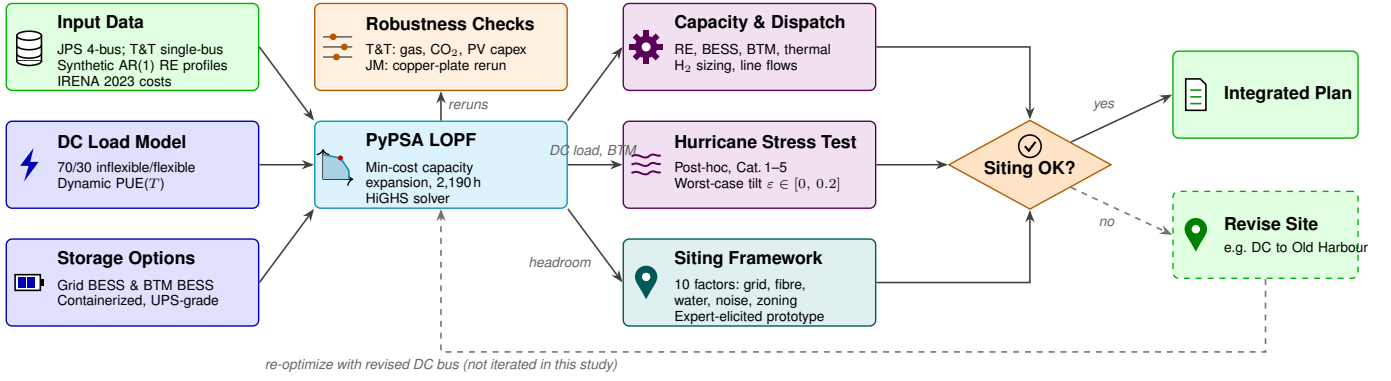
\begin{figure*}[!t]
\centering
\resizebox{\textwidth}{!}{%
\begin{tikzpicture}[
  font=\sffamily,
  box/.style={rectangle, rounded corners=3pt, draw=#1!70!black, line width=0.7pt,
              fill=#1!12, minimum width=4.5cm, minimum height=1.55cm, inner sep=3pt},
  ttl/.style={font=\sffamily\bfseries\small, anchor=west},
  sub/.style={font=\sffamily\scriptsize, anchor=north west, align=left, text width=3.6cm},
  arr/.style={-{Stealth[length=2.2mm]}, line width=0.8pt, draw=black!75},
  darr/.style={arr, dashed, draw=black!55},
  lab/.style={font=\sffamily\scriptsize\itshape, text=black!60},
  pics/db/.style={code={
    \draw[line width=0.8pt,fill=white] (-.22,-.25) rectangle (.22,.2);
    \draw[line width=0.8pt,fill=white] (0,.2) ellipse (.22 and .08);
    \draw[line width=0.8pt] (-.22,.03) arc (180:360:.22 and .08);
    \draw[line width=0.8pt] (-.22,-.12) arc (180:360:.22 and .08);
    \draw[line width=0.8pt,fill=white] (-.22,-.25) arc (180:360:.22 and .08);}},
  pics/bolt/.style={code={
    \fill[#1] (.08,.3) -- (-.16,-.02) -- (-.01,-.02) -- (-.09,-.3) -- (.17,.05) -- (.02,.05) -- cycle;}},
  pics/batt/.style={code={
    \draw[line width=0.8pt,fill=white] (-.25,-.14) rectangle (.2,.14);
    \fill (.2,-.06) rectangle (.26,.06);
    \fill[#1] (-.2,-.09) rectangle (-.07,.09); \fill[#1] (-.04,-.09) rectangle (.09,.09);}},
  pics/gear/.style={code={
    \foreach \a in {0,45,...,315}{\fill[#1,rotate=\a] (-.05,.18) rectangle (.05,.29);}
    \fill[#1] (0,0) circle (.2); \fill[white] (0,0) circle (.08);}},
  pics/lp/.style={code={
    \draw[line width=0.7pt,->] (-.26,-.26) -- (-.26,.28);
    \draw[line width=0.7pt,->] (-.26,-.26) -- (.28,-.26);
    \fill[#1!35] (-.26,.18) -- (.05,.12) -- (.2,-.1) -- (.2,-.26) -- (-.26,-.26) -- cycle;
    \draw[#1,line width=0.8pt] (-.26,.18) -- (.05,.12) -- (.2,-.1) -- (.2,-.26);
    \fill[red!75!black] (.05,.12) circle (.045);}},
  pics/storm/.style={code={
    \foreach \y in {.15,0,-.15}{\draw[#1,line width=1pt] (-.26,\y) .. controls (-.13,\y+.08) and (-.05,\y-.08) .. (.05,\y) .. controls (.13,\y+.08) and (.2,\y-.02) .. (.26,\y+.02);}}},
  pics/pin/.style={code={
    \fill[#1] (0,-.3) .. controls (-.1,-.1) and (-.2,0) .. (-.2,.1) arc (180:0:.2) .. controls (.2,0) and (.1,-.1) .. (0,-.3);
    \fill[white] (0,.1) circle (.075);}},
  pics/slider/.style={code={
    \foreach \y/\x in {.18/-.08,0/.12,-.18/-.02}{\draw[line width=0.8pt,black!60] (-.26,\y) -- (.26,\y); \fill[#1] (\x,\y) circle (.06);}}},
  pics/doc/.style={code={
    \draw[line width=0.8pt,fill=white] (-.18,-.28) -- (-.18,.28) -- (.08,.28) -- (.18,.18) -- (.18,-.28) -- cycle;
    \foreach \y in {.08,-.04,-.16}{\draw[#1,line width=0.8pt] (-.1,\y) -- (.1,\y);}}},
  pics/chk/.style={code={
    \draw[line width=0.8pt] (0,0) circle (.2);
    \draw[line width=1.1pt] (-.09,0) -- (-.02,-.08) -- (.1,.08);}},
]
\newcommand{\cell}[5]{
  \pic at ([xshift=0.42cm]#1.west) {#3=#2!70!black};
  \node[ttl] at ([xshift=0.8cm,yshift=0.42cm]#1.west) {#4};
  \node[sub] at ([xshift=0.8cm,yshift=0.18cm]#1.west) {#5};}

\node[box=green]  (inp) at (0, 2.1) {};  \cell{inp}{green}{db}{Input Data}{JPS 4-bus; T\&T single-bus\\Synthetic AR(1) RE profiles\\IRENA 2023 costs}
\node[box=blue]   (dc)  at (0, 0)   {};  \cell{dc}{blue}{bolt}{DC Load Model}{70/30 inflexible/flexible\\Dynamic PUE($T$)}
\node[box=blue]   (st)  at (0,-2.1) {};  \cell{st}{blue}{batt}{Storage Options}{Grid BESS \& BTM BESS\\Containerized, UPS-grade}
\node[box=cyan]   (lp)  at (5.5, 0)   {}; \cell{lp}{cyan}{lp}{PyPSA LOPF}{Min-cost capacity\\expansion, 2{,}190\,h\\HiGHS solver}
\node[box=orange] (rob) at (5.5, 2.1) {}; \cell{rob}{orange}{slider}{Robustness Checks}{T\&T: gas, CO$_2$, PV capex\\JM: copper-plate rerun}
\node[box=violet] (opt) at (11, 2.1) {}; \cell{opt}{violet}{gear}{Capacity \& Dispatch}{RE, BESS, BTM, thermal\\H$_2$ sizing, line flows}
\node[box=violet] (dro) at (11, 0)   {}; \cell{dro}{violet}{storm}{Hurricane Stress Test}{Post-hoc, Cat.\,1--5\\Worst-case tilt $\varepsilon\in[0,\,0.2]$}
\node[box=teal]   (sit) at (11,-2.1) {}; \cell{sit}{teal}{pin}{Siting Framework}{10 factors: grid, fibre,\\water, noise, zoning\\Expert-elicited prototype}
\node[diamond, aspect=1.6, draw=orange!70!black, line width=0.7pt, fill=orange!22,
      minimum width=2.9cm, inner sep=1pt, align=center, font=\sffamily\bfseries\small] (dec) at (16.0,0) {\\[1pt]Siting OK?};
\pic at ([yshift=0.33cm]dec.center) {chk};
\node[box=green, minimum width=3.3cm] (plan) at (20.2, 1.25) {};
\pic at ([xshift=0.42cm]plan.west) {doc=green!60!black};
\node[ttl] at ([xshift=0.8cm]plan.west) {Integrated Plan};
\node[box=green, minimum width=3.3cm, dashed] (rev) at (20.2,-1.25) {};
\pic at ([xshift=0.42cm]rev.west) {pin=green!50!black};
\node[ttl] at ([xshift=0.8cm,yshift=0.2cm]rev.west) {Revise Site};
\node[sub, text width=2.6cm] at ([xshift=0.8cm,yshift=0cm]rev.west) {e.g.\ DC to Old Harbour};

\draw[arr] (inp.east) -- (lp.north west);
\draw[arr] (dc.east)  -- (lp.west);
\draw[arr] (st.east)  -- (lp.south west);
\draw[arr] (lp.north) -- node[right,lab]{reruns} (rob.south);
\draw[arr] (lp.north east) -- (opt.west);
\draw[arr] (lp.east) -- node[above,lab,pos=.45]{DC load, BTM} (dro.west);
\draw[arr] (lp.south east) -- node[below left,lab,pos=.55]{headroom} (sit.west);
\draw[arr] (opt.east) -| (dec.north);
\draw[arr] (dro.east) -- (dec.west);
\draw[arr] (sit.east) -| (dec.south);
\draw[arr] (dec.north east) -- node[above left,lab]{yes} (plan.west);
\draw[darr] (dec.south east) -- node[below left,lab]{no} (rev.west);
\draw[darr] (rev.south) |- ([yshift=-0.45cm]sit.south) -| node[pos=0.5,below,lab]{re-optimize with revised DC bus (not iterated in this study)} (lp.south);
\end{tikzpicture}}
\caption{Integrated methodology: PyPSA capacity expansion with robustness checks, a post-hoc hurricane stress test, and multi-criteria siting evaluation. Dashed path: site-revision loop, identified but not iterated in this study.}
\label{fig:workflow}
\end{figure*}

Fig.~\ref{fig:workflow} presents the integrated methodology.

\subsection{Capacity Expansion Formulation}

The framework minimizes total annualized system cost over snapshots $\mathcal{T}$ ($|\mathcal{T}| = 2{,}190$\,h). Capital costs are scaled by $\alpha = |\mathcal{T}|/8760$ to maintain correct annualized economics for sub-annual simulation:
\begin{equation}
\min_{\substack{p_{g,t},\, \bar{G}_g \\ e_{s,t},\, \bar{E}_s}} \;\sum_{g \in \mathcal{G}} \left(\alpha\, c_g^{\text{inv}} \bar{G}_g + \sum_{t} c_g^{\text{mc}} p_{g,t}\right) + \sum_{s \in \mathcal{S}} \alpha\, c_s^{\text{inv}} \bar{E}_s
\label{eq:obj}
\end{equation}
where annualized investment cost uses the capital recovery factor:
\begin{equation}
c_g^{\text{inv}} = \text{CAPEX}_g \cdot \frac{r(1+r)^{n_g}}{(1+r)^{n_g}-1} + \text{OPEX}_g
\label{eq:crf}
\end{equation}
with weighted average cost of capital $r$ and asset lifetime $n_g$.

\noindent\textbf{Nodal power balance} at each bus~$b$ and snapshot~$t$:
\begin{equation}
\sum_{g \in \mathcal{G}_b} p_{g,t} + \sum_{l \in \mathcal{L}_b^{+}} \eta_l f_{l,t} - \sum_{l \in \mathcal{L}_b^{-}} f_{l,t} = d_{b,t}^{\text{grid}} + d_{b,t}^{\text{DC}} \quad \forall\, b,t
\label{eq:balance}
\end{equation}
where $f_{l,t}$ is the power flow on link~$l$ and $\eta_l$ is its efficiency (capturing ohmic losses on the 138\,kV backbone: $\eta_{\text{KGN-OH}}=0.990$, $\eta_{\text{OH-MP}}=0.985$, $\eta_{\text{MP-MB}}=0.975$).

\noindent\textbf{Generator bounds} enforce minimum stable generation and time-varying capacity factors for variable RE:
\begin{equation}
\underline{p}_g (\bar{G}_g^{0} + \bar{G}_g) \leq p_{g,t} \leq \bar{p}_{g,t}\, (\bar{G}_g^{0} + \bar{G}_g) \quad \forall\, g,t
\label{eq:gen}
\end{equation}
where $\bar{G}_g^{0}$ is existing (fixed) capacity, $\bar{G}_g$ is optimized new capacity, $\underline{p}_g$ is the minimum stable output fraction (e.g., 0.40 for JEP LNG, 0.35 for HFO), and $\bar{p}_{g,t} \in [0,1]$ is the capacity factor profile for solar and wind generators.

\noindent\textbf{Transmission flow bounds}:
\begin{equation}
0 \leq f_{l,t} \leq \bar{F}_l \quad \forall\, l,t
\label{eq:tx}
\end{equation}
where $\bar{F}_l$ is the thermal rating (300/200/180\,MW for the three Jamaica corridors).

\noindent\textbf{Store energy balance} with cyclic boundary condition:
\begin{equation}
e_{s,t} = e_{s,t-1} + \eta_s^{+} p_{s,t}^{+} - \frac{p_{s,t}^{-}}{\eta_s^{-}}, \quad e_{s,0} = e_{s,|\mathcal{T}|}
\label{eq:store}
\end{equation}
where $\eta_s^{+}$, $\eta_s^{-}$ are charge/discharge efficiencies. Grid-scale BESS uses $\eta_{\text{RT}} = 0.865$ (round-trip); BTM BESS uses $\eta_{\text{RT}} = 0.846$ reflecting tropical derate and containerized power electronics.

\subsection{Data Center Load Model}

DC electrical demand decomposes into inflexible (latency-sensitive) and flexible (batch/training) components:
\begin{equation}
d_t^{\text{DC}} = \underbrace{(1-\phi)\, P_t^{\text{IT}}\, \text{PUE}(T_t)}_{\text{inflexible}} + \underbrace{\phi\, \overline{P^{\text{IT}} \cdot \text{PUE}}^{\,\text{day}}}_{\text{flexible (daily avg.)}}
\label{eq:dc}
\end{equation}
with flexible fraction $\phi=0.30$ and dynamic PUE responding to ambient temperature:
\begin{equation}
\text{PUE}(T_t) = \text{PUE}_{\text{nom}}\bigl(1 + 0.005\,(T_t - 25\,^\circ\text{C})\bigr)
\label{eq:pue}
\end{equation}
The IT load $P_t^{\text{IT}}$ exhibits diurnal variation ($\pm 6\%$, peak at 14:00) and weekday/weekend modulation (12\% reduction on weekends). The flexible portion is modeled as a daily-averaged flat load, representing the optimizer's freedom to shift batch workloads within a 24-hour window without introducing additional integer or store-based decision variables.

\subsection{Behind-the-Meter BESS}

BTM storage is modeled as a separate store attached to the bus that serves the data center (the Kingston bus in Case~B, the single bus in Case~A). It shares that bus with any grid-scale BESS and differs only in its parameters: a 19\% energy cost premium (\$250 vs.\ \$210/kWh) for containerized UPS-grade packaging, a shorter life (12 vs.\ 15 years), and $\eta=0.92$ per direction (vs.\ 0.93). The LP therefore treats BTM as a higher-cost battery at the same node and builds it only once the grid-BESS build limit binds. Its islanding value is assessed only in the stress test (Section~III-D); a dedicated DC bus behind a rated feeder is future work.

\subsection{Post-Hoc DRO Resilience}

Hurricane resilience is assessed outside the optimization as a stylized distributionally robust stress test. For island~$i$, each Saffir--Simpson category $c\in\{1,\dots,5\}$ has a damage profile $(p_c, g_c, h_c)$: base share of hurricanes $p_c$, fractional grid outage $g_c$, and outage duration $h_c$\,(h). The annual event rate is $\lambda_{i,c}=s_i p_c$, where $s_i$ is an island exposure multiplier. Unserved DC energy per event is
\begin{equation}
\text{UE}_{i,c}=\bigl[P_i^{\text{DC}}h_c-0.5(1-g_c)P_i^{\text{DC}}h_c-E_i^{\text{BTM}}(1-0.3g_c)\bigr]^{+}
\label{eq:ue}
\end{equation}
assuming 50\% residual grid supply on non-outaged feeders and BTM availability derated by $0.3g_c$. Resilience is
\begin{equation}
R_i(\varepsilon) = 1 - \frac{\sum_c \tilde{\lambda}_{i,c}(\varepsilon)\,\text{UE}_{i,c}}{P_i^{\text{DC}} \cdot 8760},\quad \tilde{\lambda}_{i,c}=\bigl[s_i p_c+\varepsilon(2g_c-0.5)\bigr]_{0.001}^{0.999}
\label{eq:dro}
\end{equation}
The adversarial tilt moves frequency from low-outage toward high-outage categories as $\varepsilon$ grows. It is a tractable surrogate for the worst case over a Wasserstein ball on the category distribution, not an exact DRO solution.

\textit{Parameter basis}: $p_c=(0.35,0.25,0.20,0.15,0.05)$ is set to be broadly consistent with the peak-intensity distribution of Atlantic hurricanes in NOAA HURDAT2~\cite{hurdat2}. $g_c$ (0.15--0.90) and $h_c$ (24--240\,h) are engineering judgement, not fitted to utility outage records. $s_{\text{Trinidad}}=0.3$ and $s_{\text{Jamaica}}=0.8$ reflect Trinidad's position at the southern margin of the Atlantic hurricane track (${\approx}10.5^\circ$N) relative to Jamaica. The radius range $\varepsilon\in[0,0.20]$ is a sensitivity sweep rather than an estimate; it represents a plausible shift toward intense storms, consistent in direction with projected increases in the proportion of Category~4--5 cyclones~\cite{ipcc_ar6}. $E^{\text{BTM}}$ equals the modeled BTM capacity (30/84\,MWh). All values are illustrative; calibration against JPS and T\&TEC restoration records is future work.

\section{Case Studies}

\begin{figure}[htbp]
\centering
\includegraphics[width=\columnwidth]{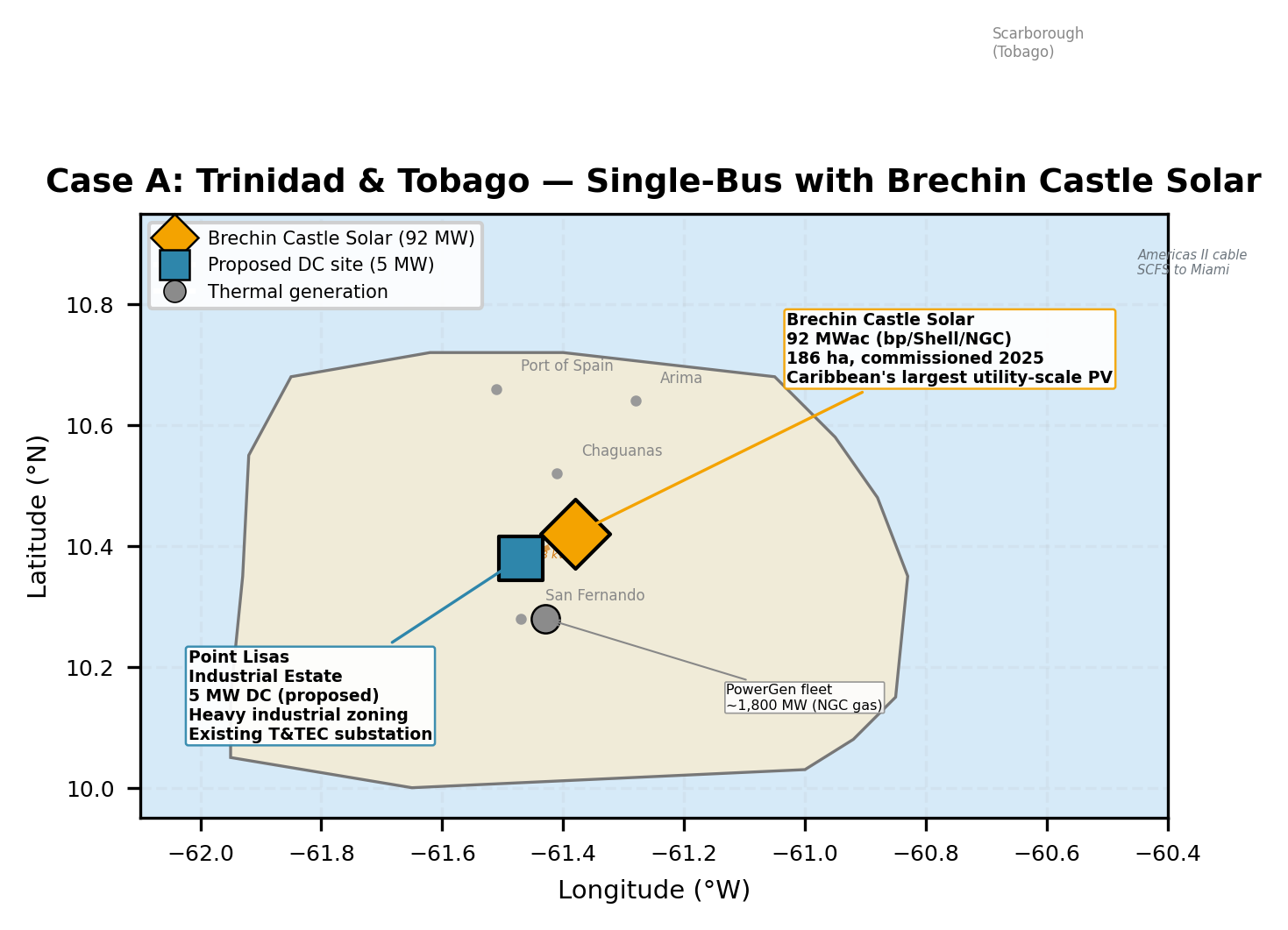}
\caption{Case A: Trinidad \& Tobago. Brechin Castle Solar (92\,MWac) and proposed 5\,MW data center at Point Lisas Industrial Estate.}
\label{fig:trinidad}
\end{figure}

\subsection{Case A: Trinidad \& Tobago (Fig.~\ref{fig:trinidad})}

Single-bus: 1{,}800\,MW gas fleet, Brechin Castle 92\,MWac solar~(bp/Shell/NGC joint venture, commissioned 2025, the Caribbean's largest utility-scale PV~\cite{brechin2025}), extendable solar (300\,MW), wind (100\,MW), grid BESS (400\,MWh), H$_2$ (20\,MW). WACC~7\%. DC: 5\,MW~IT at Point Lisas Industrial Estate, PUE~1.5, 30\%~flexible, 30\,MWh BTM.

Trinidad~\&~Tobago's energy landscape is evolving: while natural gas remains the backbone of electricity generation, the Brechin Castle project signals a strategic diversification pathway. The model captures both the existing gas fleet and the new solar capacity, reflecting the nation's current energy mix without prejudging future policy directions.

\begin{figure}[htbp]
\centering
\includegraphics[width=\columnwidth]{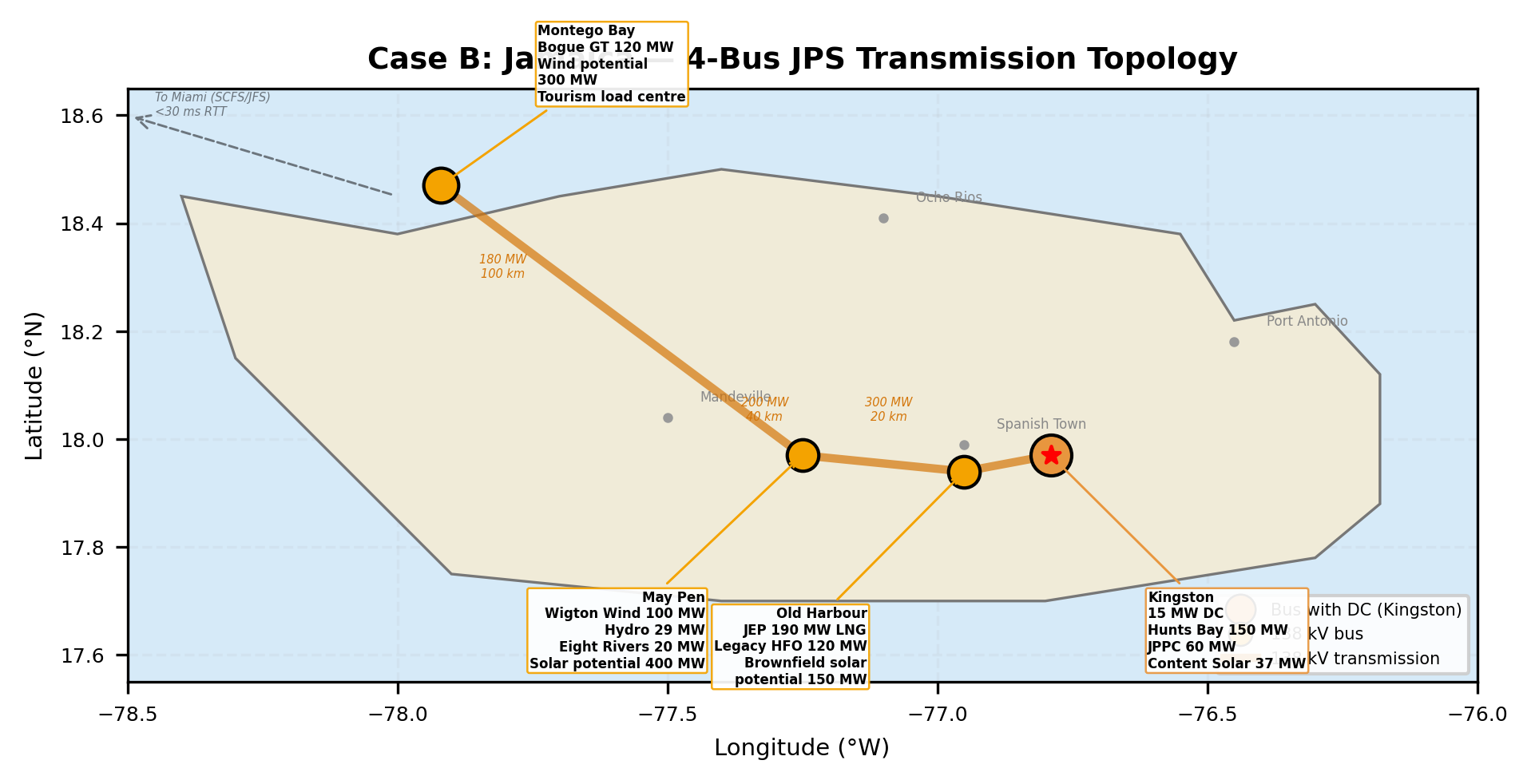}
\caption{Case B: Jamaica 4-bus topology based on JPS 138\,kV backbone. Buses correspond to real substations. Red star marks the 15\,MW DC at Kingston.}
\label{fig:jamaica}
\end{figure}

\subsection{Case B: Jamaica (Fig.~\ref{fig:jamaica})}

Four buses model the JPS 138\,kV backbone: \textbf{Kingston}~(load centre, Hunts Bay 150\,MW, JPPC/Rockfort 120\,MW diesel, Content Solar 37\,MW, \emph{15\,MW DC}); \textbf{Old Harbour}~(JEP 190\,MW LNG combined-cycle~\cite{jep2019}, 120\,MW legacy HFO, brownfield solar 150\,MW); \textbf{May Pen}~(Wigton Wind 100\,MW~\cite{wigton}, 29\,MW hydro, Eight Rivers Solar 20\,MW, extendable solar 400\,MW); \textbf{Montego Bay}~(Bogue 120\,MW GT, extendable wind 300\,MW). Transmission: KGN--OH 300\,MW (20\,km, recently upgraded~\cite{jps_upgrade}), OH--MP 200\,MW (40\,km), MP--MB 180\,MW (100\,km). WACC~10\%.

\begin{table}[htbp]
\centering
\caption{Technology Cost Assumptions (IRENA 2023~\cite{irena2023}, Caribbean premiums applied)}
\label{tab:costs}
\setlength{\tabcolsep}{3pt}
\begin{tabular}{@{}lcccc@{}}
\toprule
Technology & CAPEX & OPEX & Life & Source \\
\midrule
Solar PV & \$1{,}100/kW & \$12/kW/yr & 25\,yr & \cite{irena2023} \\
Onshore wind & \$1{,}600/kW & \$35/kW/yr & 20\,yr & \cite{irena2023} \\
Grid BESS (4h) & \$210/kWh & \$4/kWh/yr & 15\,yr & \cite{irena2023} \\
BTM BESS (4h) & \$250/kWh & \$5/kWh/yr & 12\,yr & Est.\ (+19\%) \\
PEM electrolyzer & \$1{,}400/kW & \$28/kW/yr & 20\,yr & \cite{irena2023} \\
\bottomrule
\end{tabular}
\end{table}

\noindent All profiles use synthetic AR(1)-modulated solar/wind capacity factors calibrated to Caribbean averages (solar CF\,0.18--0.21, wind CF\,0.22--0.30). The simulation covers Q1 (2{,}190\,h) with investment costs scaled by $\alpha = 2190/8760$.

\section{Results}

\begin{figure}[htbp]
\centering
\includegraphics[width=\columnwidth]{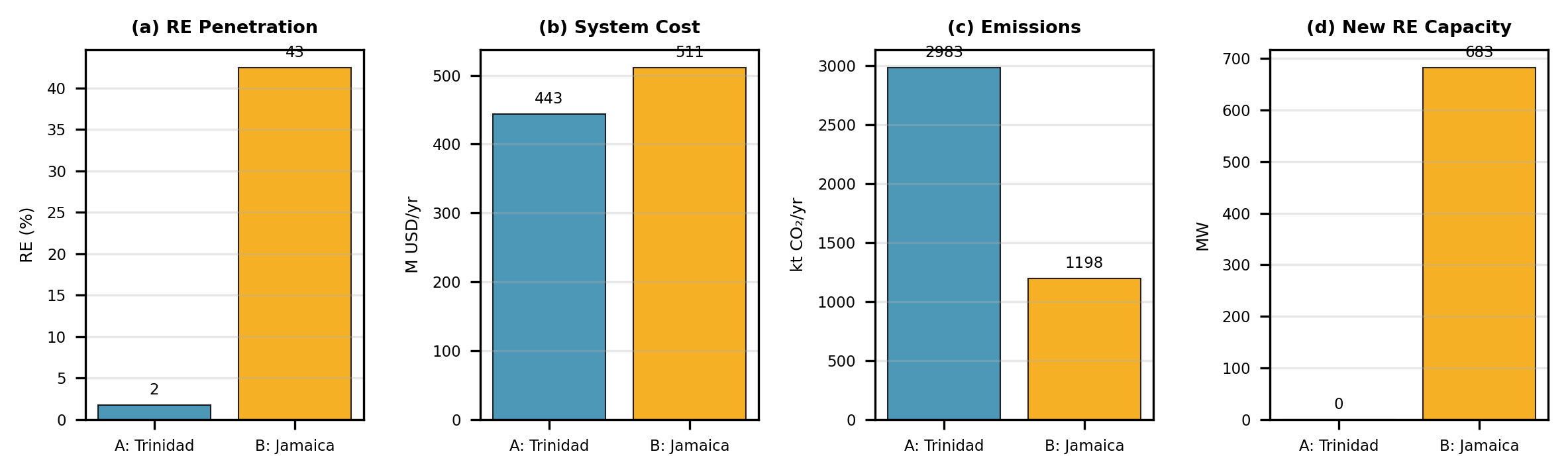}
\caption{Optimization results: (a)~RE penetration, (b)~annualized system cost, (c)~CO$_2$ emissions, (d)~total new RE capacity.}
\label{fig:results}
\end{figure}

Table~\ref{tab:results} and Fig.~\ref{fig:results} summarize the optimized outcomes.

\begin{table}[htbp]
\centering
\caption{Optimization Results (Annualized)}
\label{tab:results}
\begin{tabular}{@{}lcc@{}}
\toprule
Metric & Case A: Trinidad & Case B: Jamaica \\
\midrule
System cost (M\,USD/yr) & 443 & 511 \\
RE fraction (\%) & 1.8 & 42.5 \\
New solar (MW) & 0 & 405 \\
New wind (MW) & 0 & 278 \\
Grid BESS (MWh) & 0 & 800 \\
BTM BESS (MWh) & 0 & 84 \\
CO$_2$ (kt/yr) & 2{,}983 & 1{,}198 \\
\bottomrule
\end{tabular}
\end{table}

\textbf{Trinidad}: The optimizer dispatches Brechin Castle's 92\,MW (1.8\% RE fraction) but builds no additional RE or storage. At a gas cost of \$55/MWh, new solar (levelized ${\approx}$\$67/MWh at 7\% WACC and CF\,0.18) is not competitive.

\textit{Robustness}: Table~\ref{tab:sens} varies gas cost, carbon price, and solar CAPEX one at a time. The response is a threshold. Solar enters between \$65 and \$70/MWh gas cost, at a carbon price between \$25 and \$50/tCO$_2$ (equivalent to a \$9--19/MWh adder at 0.37\,tCO$_2$/MWh), or at \$800/kW solar CAPEX. When it enters, all 300\,MW of candidate solar is built, and wind follows at \$100/MWh. RE rises only to 7.5--9.9\% and CO$_2$ falls by 6--8\%. Storage and BTM never deploy in the tested range. Two features limit uptake: the candidate capacity caps and the 25\% minimum stable output of the 1{,}800\,MW gas fleet. The zero-RE result is therefore robust at current gas costs but sensitive to modest carbon pricing or cost declines.

\begin{table}[htbp]
\centering
\caption{Case A Sensitivity (one-at-a-time; base: gas \$55/MWh, solar \$1{,}100/kW, no carbon price)}
\label{tab:sens}
\setlength{\tabcolsep}{3pt}
\begin{tabular}{@{}lccccc@{}}
\toprule
Scenario & New PV & New wind & RE & Cost & CO$_2$ \\
 & (MW) & (MW) & (\%) & (M\$/yr) & (kt/yr) \\
\midrule
Base & 0 & 0 & 1.8 & 443 & 2{,}983 \\
Gas \$45--65/MWh & 0 & 0 & 1.8 & 363--524 & 2{,}983 \\
Gas \$70/MWh & 300 & 0 & 7.5 & 563 & 2{,}808 \\
Gas \$85/MWh & 300 & 0 & 7.5 & 677 & 2{,}808 \\
Gas \$100/MWh & 300 & 100 & 9.9 & 791 & 2{,}737 \\
CO$_2$ \$25/t & 0 & 0 & 1.8 & 518 & 2{,}983 \\
CO$_2$ \$50/t & 300 & 0 & 7.5 & 590 & 2{,}808 \\
Solar \$800/kW & 300 & 0 & 7.5 & 442 & 2{,}808 \\
\bottomrule
\end{tabular}
\end{table}

\textbf{Jamaica}: The optimizer deploys 405\,MW solar across May Pen and Old Harbour and 278\,MW wind at Montego Bay. Grid BESS reaches its 800\,MWh build limit at Kingston and Old Harbour, and 84\,MWh BTM BESS deploys at the Kingston bus, also at its limit. BTM is chosen despite its cost premium because the grid-BESS limits bind. A copper-plate rerun (unconstrained, lossless transmission) deploys the same storage (800\,MWh grid, 84\,MWh BTM), 400\,MW solar, and 281\,MW wind, at \$507M/yr (0.8\% lower). The 4-bus network is congested for only 54\,h (MP$\rightarrow$OH) and 13\,h (OH$\rightarrow$KGN) of 2{,}190\,h. Storage deployment is therefore driven by solar variability and the storage build limits rather than by transmission congestion. The 42.5\% Q1 RE share is consistent with Jamaica's 50\%-by-2030 target.

\textbf{DRO Resilience}: Trinidad remains at $R \geq 0.989$ ($0.998\rightarrow0.989$) across $\varepsilon\in[0,0.20]$, reflecting its low exposure multiplier. Jamaica degrades from $R = 0.996$ to $R = 0.987$ at $\varepsilon = 0.20$, driven by higher strike exposure. In both cases the BTM provides about 4~hours of islanded DC backup at full load (84\,MWh for a 21\,MW facility load in Jamaica). These values inherit the illustrative parameters of Section~III-D and should be read comparatively, not as absolute availability estimates.

\section{Multi-Criteria Siting Framework}

PyPSA determines \emph{where to build generation}; it does not assess whether a data center is \emph{socially and regulatorily feasible} at a given site. Caribbean SIDS lack the zoning frameworks that US jurisdictions have adopted in response to community opposition~\cite{dcwatch2025,stafford2025,albemarle2025,ramboll2024}. We propose a 10-factor framework for Caribbean conditions (Table~\ref{tab:siting}). The current version is an \emph{expert-elicited prototype}. The authors assigned scores against the anchors in Table~\ref{tab:rubric}, using the LP results (grid headroom, transmission), the cost and resource assumptions of Section~IV (RE quality), and public infrastructure and hazard information. Scores are unweighted and have not yet been validated with stakeholders; formal MCDA is planned (Section~VII).

\begin{table}[htbp]
\centering
\caption{Scoring Anchors for Table~\ref{tab:siting} (Prototype)}
\label{tab:rubric}
\scriptsize
\setlength{\tabcolsep}{2pt}
\begin{tabular}{@{}p{1.9cm}p{3.1cm}p{3.1cm}@{}}
\toprule
Factor & Score 5 & Score 1 \\
\midrule
Grid headroom & Large reserve margin; DC ${<}1\%$ of peak & Tight margin; DC ${>}5\%$ of peak \\
RE resource & Solar CF ${\geq}0.21$ or wind CF ${\geq}0.30$ nearby & Weak resource, little buildable land \\
Transmission & HV substation on site; no congestion & New HV line or congested corridor \\
Fibre RTT & ${<}30$\,ms to Miami, diverse cables & ${>}60$\,ms or single cable \\
IX/peering & IXP in same metro & No domestic IXP nearby \\
Hurricane & Rarely on track; outside surge zone & Frequent major-storm track; coastal \\
Water & Unstressed or non-potable source & Stressed municipal supply only \\
Seismic & Low hazard zone & High hazard; near active fault \\
Noise buffer & ${\geq}150$\,m to residences feasible & Residences ${<}50$\,m \\
Zoning & Heavy-industrial zoning permits use & Rezoning and full EIA needed \\
\bottomrule
\end{tabular}
\end{table}

\begin{table}[htbp]
\centering
\caption{DC Siting Suitability Framework (1--5 scale, 5\,=\,best)}
\label{tab:siting}
\setlength{\tabcolsep}{2.5pt}
\begin{tabular}{@{}p{3.0cm}ccccc@{}}
\toprule
Factor & TT & JM-KGN & JM-OH & JM-MB \\
\midrule
\multicolumn{5}{@{}l}{\textit{Energy \& Infrastructure}} \\
\;\; Grid headroom & 5 & 3 & 4 & 2 \\
\;\; RE resource quality & 3 & 3 & 4 & 5 \\
\;\; Transmission adequacy & 5 & 4 & 4 & 2 \\
\multicolumn{5}{@{}l}{\textit{Connectivity}} \\
\;\; Submarine fibre RTT & 4 & 5 & 3 & 3 \\
\;\; IX/peering proximity & 3 & 4 & 1 & 2 \\
\multicolumn{5}{@{}l}{\textit{Environmental \& Climate}} \\
\;\; Hurricane exposure & 5 & 3 & 3 & 2 \\
\;\; Water availability & 4 & 3 & 4 & 3 \\
\;\; Seismic risk & 4 & 3 & 3 & 3 \\
\multicolumn{5}{@{}l}{\textit{Societal \& Regulatory}} \\
\;\; Noise buffer feasibility & 4 & 2 & 4 & 3 \\
\;\; Zoning readiness & 4 & 3 & 3 & 2 \\
\midrule
\textbf{Total (unweighted)} & \textbf{41} & \textbf{33} & \textbf{33} & \textbf{27} \\
\bottomrule
\end{tabular}
\end{table}

\textbf{Key finding}: Old Harbour---not Kingston---emerges as Jamaica's strongest DC site (tied at 33, but with superior noise/water/grid attributes). This is a brownfield industrial location near the JEP LNG plant with existing 138\,kV connectivity. The energy optimization placed the DC at Kingston for latency; the siting framework reveals Old Harbour may be more feasible, illustrating the tension this framework is designed to surface.

\section{Critical Assessment}

We identify six categories of weakness in this work and suggest paths for future improvement.

\textit{1) Synthetic profiles}: All RE capacity factors and load profiles are generated from AR(1) stochastic models calibrated to Caribbean averages rather than measured meteorological or SCADA data. This introduces unquantified bias in temporal correlations. Future work should incorporate ERA5 reanalysis data or locally measured irradiance/wind speed time series.

\textit{2) Sub-annual simulation}: The Q1 snapshot (2{,}190\,h) captures dry-season trade-wind patterns but underrepresents hurricane season (Aug--Nov) and wet-season hydro variability. The scaling factor $\alpha = 2190/8760$ assumes Q1 is representative of annualized economics, which may overvalue dry-season wind and undervalue storage needed for hurricane resilience. Full 8{,}760-hour simulation---or representative period selection via k-medoids clustering---would improve robustness.

\textit{3) Simplified workload flexibility}: The daily-average flattening model understates the value of hourly temporal shifting. A full formulation using store-based shift operators~\cite{google247} or explicit bin-packing constraints would better capture RE--DC synergies, at the cost of significantly increased LP complexity.

\textit{4) Missing economic dimensions}: The LP captures energy arbitrage but not demand charges, capacity markets, or ancillary service revenues that drive BTM deployment in practice. Carbon pricing is examined only as a one-at-a-time sensitivity for Trinidad (Table~\ref{tab:sens}); no emission constraint is modeled. Sector coupling (waste heat recovery via absorption chillers, COP~$\approx$\,0.7) could offset 15--20\% of cooling electricity but adds multi-carrier complexity.

\textit{5) Siting score subjectivity}: The 10-factor framework uses expert-assigned scores guided by qualitative anchors (Table~\ref{tab:rubric}) rather than quantitative indices. Weighting is uniform, yet factors like submarine cable latency may dominate for certain workload profiles. A formal multi-criteria decision analysis (MCDA) with stakeholder-elicited weights~\cite{uli2024} and sensitivity analysis over weight vectors would strengthen the framework's policy relevance.

\textit{6) Network simplification}: The 4-bus Jamaica model captures major transmission corridors but omits the 69\,kV sub-transmission network, reactive power constraints, and N-1 contingency analysis required by the JPS Transmission Code~\cite{jps_grid}. More detailed representations using PyPSA-Earth workflows could address this gap.

\section{Discussion}

\textit{Why this work matters for Caribbean SIDS}: We are not aware of a published study that applies open-source capacity expansion modeling to data center co-planning in the Caribbean. Existing tools address energy economics (HOMER, PLEXOS) or real estate (CBRE, JLL) separately. For SIDS policymakers assessing data center proposals, which often arrive as unsolicited foreign investment, the framework offers a transparent, license-free assessment.

\textit{What the network adds}: In Jamaica, RE concentrates at May Pen and Montego Bay while the DC load sits at Kingston. At the present 138\,kV ratings, however, the copper-plate comparison changes cost by only 0.8\% and leaves storage unchanged. The multi-bus model is still necessary to confirm this, and it becomes decisive under larger DC loads or corridor outages (N-1). These cases are the natural next test for IRP-type studies.

\textit{The Brechin Castle model for SIDS}: The bp/Shell/NGC consortium behind Brechin Castle pairs international developers with a national gas company under government facilitation. This structure addresses the financing and capacity constraints facing utility-scale RE in smaller SIDS~\cite{brechin_ecp}. A Point Lisas data center could combine existing industrial zoning with the adjacent solar. Table~\ref{tab:sens}, however, shows that additional RE there depends on gas pricing or carbon policy.

\textit{Societal factors dominate SIDS siting}: A 500-foot setback, now adopted in parts of Virginia~\cite{stafford2025}, could span an entire coastal settlement in Dominica or St.~Kitts. Even in Jamaica, industrial land is confined to a few corridors. Point Lisas and Old Harbour are rare brownfield sites with heavy-industrial zoning and room for buffers, and they should be prioritized over greenfield sites that require new zoning.

\textit{Water as a binding constraint}: A 15\,MW data center using evaporative cooling consumes approximately 50\,ML/yr~\cite{compass2025}. On water-stressed islands this can trigger the community opposition seen in the US~\cite{dcwatch2025}. Closed-loop or air-cooled alternatives add 15--25\% to cooling CAPEX, a premium not yet captured in the optimization.

\textit{Policy recommendations}: Caribbean SIDS should (i)~develop data-center-specific zoning addressing noise (45--55\,dBA limits), water usage caps, and setback distances scaled to island geography; (ii)~require co-located RE and BTM storage as permit conditions; and (iii)~prioritize brownfield industrial sites where infrastructure and buffer zones exist.

\section{Conclusion}

This paper presented a PyPSA framework for DC--RE co-planning in Caribbean SIDS. Jamaica's 4-bus model, grounded in JPS topology, reaches 42.5\% RE with 800\,MWh grid BESS and 84\,MWh BTM BESS. A copper-plate comparison shows that storage deployment there is driven by storage build limits rather than congestion. For Trinidad, sensitivity analysis shows the zero-new-RE optimum reverses at moderate carbon prices (\$25--50/tCO$_2$) or solar CAPEX near \$800/kW. The 10-factor siting framework reveals that Old Harbour, not Kingston, may be Jamaica's most suitable DC location when societal factors are considered. For Trinidad, the Brechin Castle solar project---a consortium of bp, Shell, and NGC delivering 92\,MWac adjacent to the Point Lisas industrial corridor---demonstrates a replicable pathway for RE integration that future DC developments could directly leverage.

The key message is that energy optimization alone is insufficient: societal impact, regulatory readiness, and climate resilience must be co-evaluated. As data center demand reaches Caribbean shores, integrated tools combining open-source capacity optimization with multi-criteria siting assessment will be essential for ensuring these developments serve the interests of small island communities.

\textbf{Future work} should pursue four directions: (i)~full 8{,}760-hour simulation with ERA5 reanalysis-derived profiles to eliminate seasonal bias; (ii)~formal MCDA with stakeholder-elicited weights for the siting framework; (iii)~extension to additional SIDS contexts (Barbados, Eastern Caribbean states, Pacific SIDS) where grid characteristics differ; and (iv)~calibration of hurricane damage parameters to utility outage records, and full carbon pricing and NDC emission constraints into the PyPSA formulation to capture the full policy landscape of Caribbean energy transition.

\section*{Code Availability and Acknowledgments}

Code, inputs, results, and figures: \url{https://github.com/sramharack/caribcon26-ocreds}. The authors thank the PyPSA~\cite{pypsa} and HiGHS~\cite{highs} developers.

\balance

\end{document}